\documentclass[twocolumn,amsmath,amssymb,prb]{revtex4}
\usepackage{graphicx}
\usepackage{hyperref}
\begin{document}
\title{Tangent planes and the mean-field approximation}
\author{David Liao}
\affiliation{Harvey Mudd College, Claremont, CA, 91711}
\homepage{http://odin.ac.hmc.edu/~dliao/}
\date{\today}
\begin{abstract}
Local linearization is highlighted to explain the success of the orbital approximation in positive response to Scerri's comments on the electronic configuration model.  The relevance of Rydberg states is made clear.  
\end{abstract}
\pacs{Empty}
\maketitle
\section{Introduction}
Eric Scerri previously highlighted the importance of understanding the mechanisms that enable the electronic configuration model to approximate atomic and molecular states, often with remarkable success.\cite{Scerri1991}  Furthermore, Scerri explained that this model cannot be explained by identifying regimes in which the interelectronic repulsion vanishes, at least not without detaching theory from empirical results.   This author agrees, and the following notes will assist instructors to present orbitals in a manner sensitive to both of these points.  

\section{Tangent plane}
A popular approach to orbitals in the mathematics community draws on the tangent plane approximation.  A guess-and-check approach illustrates.  One guesses that the solution to the $N$-electron atomic Hamiltonian
\begin{equation}\label{E:Atomic}
\left(\sum_{i=1}^N-\frac{1}{2}\nabla_i^2 + \sum_{i=1}^N-\frac{Z}{r_i} + \sum_{i>j}\frac{1}{r_{ij}}\right)\psi = E\psi
\end{equation}
roughly separates into a Hartree product\footnote{I will ignore permutation symmetry in this note.  }
\begin{equation}\label{E:HartreeProduct}
\psi \approx \prod_{i=1}^{N} \phi_i(\mathbf{r_i})
\end{equation}
with spherically symmetric orbitals, where each orbital $\phi_i$ possesses a characteristic radius.\footnote{Students of introductory quantum chemistry will recall that orbitals can have angular variation, so they are not strictly spherically symmetric.  }  This means that each pair of electrons is expected to form a right angle such that a set of Cartesian axes $x$ and $y$ can intercept the electrons with the nucleus intercepting the origin.\footnote{Even in antisymmetrized wavefunctions, bi-electronic angles are close to $90^{\circ}$.\cite{Koga2003}}

Because the electrons reside near reference radii $r_i^0$ and $r_j^0$, the repulsive potential $V_{i,j}^{REP} = 1/r_{ij}$ can be approximated by a tangent plane\cite[243]{Colley2002}
\begin{equation}\label{E:tangentplane}
V_{i,j}^{REP} \approx V_{i,j}^{TANGENT} = A_{i,j}V_i(r_i) + A_{j,i}V_j(r_j)
\end{equation}
using coefficients denoted commonly\cite{DiRocco2002} as $A_{i,j}$ and $A_{j,i}$ and a two-dimensional coordinate system $\left(V_i(r_i), V_j(r_j)\right)$ obtained from $\left(r_i,r_j\right)$.\footnote{Three variables are required to describe the spherically-symmetric ground-state of a two-electron system.  In fact, the Taylor expansion of $1/r_{ij} = \left(r_i^2 + r_j^2 - 2r_ir_j\cos{\theta_{ij}}\right)^{-1/2}$ in variables $1/r_i$, $1/r_j$, \textit{and} $\cos{\theta_{ij}}$ is $\left[\left(\frac{r_i^0}{r_{ij}^0}\right)^3\left(1 - \frac{r_j^0}{r_i^0}\cos{\theta_{ij}^0}\right)\right]\frac{1}{r_i} + \left[\left(\frac{r_j^0}{r_{ij}^0}\right)^3\left(1 - \frac{r_i^0}{r_j^0}\cos{\theta_{ij}^0}\right)\right]\frac{1}{r_j} + \frac{r_i^0r_j^0}{\left(r_{ij}^0\right)^3}\left(\cos{\theta_{ij}}-\cos{\theta_{ij}^0}\right) + \mathcal{O}(\epsilon^2)$, hence equation \ref{E:tangentplane} drops the first-order term in the cosine of the bi-electronic angle.\cite{DiRocco2002}  Diverse coordinate systems enable Taylor expansions, but the transformation $r_i \rightarrow 1/r_i$ is chosen since it can reproduce the exact repulsive potential along any line with fixed bi-electronic angle $\theta_{ij}$ and proportionality $r_i = \alpha r_j$ between electronic radii.  }  Mathematicians refer to the tangent plane as a local linearization to draw an analogy with the tangent line approximation in one-dimensional calculus.  

\begin{figure}[htb]
\includegraphics[height=3in,width=3in,angle=0]{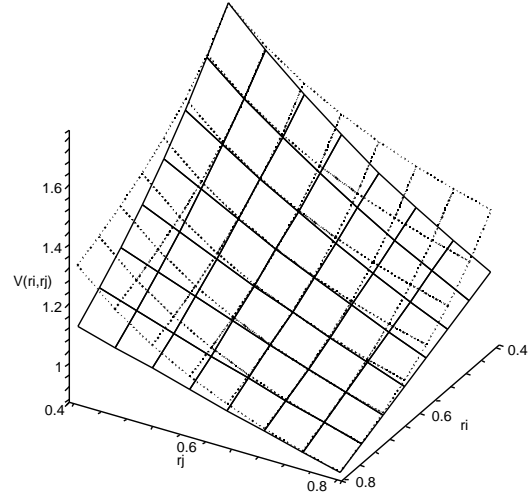}
\caption{The repulsive potential energy between two electrons on Cartesian axes $x$ and $y$ at radii $r_i$ and $r_j$ is evaluated exactly (solid lines) and by tangent plane (dash-dot lines).  }\label{F:CompareTaylor}
\end{figure}

The transformation and coefficients
\begin{align}\label{E:ScreenedTransformation}
V_i(r_i) = \frac{1}{r_i},~V_j(r_j) = \frac{1}{r_j} \\
A_{i,j} = A_{j,i} = \frac{1}{2\sqrt{2}}
\end{align}
for instance, approximate the repulsive potential when electrons lie on Cartesian axes at similar radii, as figure \ref{F:CompareTaylor} shows.  

\begin{figure}[htb]
\includegraphics[height=3in,width=3in,angle=0]{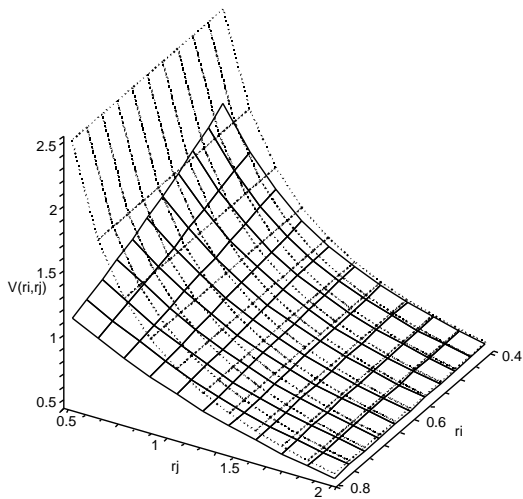}
\caption{A tangent ``plane'' (dash-dot lines) successfully approximates the repulsive potential (solid lines) between two electrons over a wide range of electronic positions $r_i$ and $r_j$ relevant in Rydberg states.  The electron at distance $r_j$ is the ``outer'' electron.  }\label{F:CompareRydberg}
\end{figure}

Local linearization is particularly appropriate for Rydberg states, where one electron is far removed from the core.  Slater recognized that a function with coefficients
\begin{align}\label{E:ScreenedRydberg}
A_{i,j} = 0\\
A_{j,i} = 1
\end{align}
along with the transformation in equation \ref{E:ScreenedTransformation}, approximates the exact potential, as illustrated in figure \ref{F:CompareRydberg}.\cite{Slater1928}  The accuracy of this tangent function in Rydberg systems has invited high-precision studies of QED corrections to non-relativistic quantum mechanics.\cite{Drake1993}

The tangent ``plane'' $V_{i,j}^{TANGENT}$ is separable, so eigenfunctions look locally like Hartree-products.  The guess-and-check demonstration is complete.  

Local linearization utilizing the coordinate system in equation \ref{E:ScreenedTransformation} provides convenient model calculations in atomic systems.  Kregar and others have calculated coefficients $A_{i,j}$ and atomic energies with programs appropriate for undergraduate coursework, especially in courses touching on Slater's rules.\cite[See the references in Di Rocco's paper.  ]{DiRocco2002,SlatersRules}  Most production-quality wavefunction packages, however, use many-body perturbation theory, coupled-cluster, or $r12$ techniques.  The primary utility of local linearization, then, is to provide intuition for understanding the suitability and limits of orbitals as a first step in high-level calculations.  

\section{Conclusion}
This letter highlights the method of local linearization, which communicates a reason for orbitals to succeed while at the same time emphasizing their approximate nature.  The hydrogenic potentials in equation \ref{E:ScreenedTransformation} invite numerical work in introductory courses.  Of course, instructors have many options for introducing students to orbitals.  Derivations of asymptotic error bounds of mean-field energies in the large atom, $Z\rightarrow \infty$, limit might interest students with advanced mathematics backgrounds.\cite{Bach1992}  All concepts should be evaluated in terms of pedagogical benefits and costs, and instructors and students working together will ultimately decide which concepts are appropriate for specific lesson plans.  

\begin{acknowledgments}
The author thanks Professor Robert Cave for his interest and support of this research.  
\end{acknowledgments}

\bibliographystyle{unsrt}
\bibliography{quantumchem.2004-10-16}

\end{document}